\documentclass[sigconf,nonacm]{acmart}

\AtBeginDocument{%
  \providecommand\BibTeX{{%
    \normalfont B\kern-0.5em{\scshape i\kern-0.25em b}\kern-0.8em\TeX}}}

\copyrightyear{2023} 
\acmYear{2023} 
\setcopyright{rightsretained} 
\acmConference[WDC '23]{The 2nd Workshop on the security implications of Deepfakes and Cheapfakes}{July 10, 2023}{Melbourne, VIC, Australia}
\acmBooktitle{The 2nd Workshop on the security implications of Deepfakes and Cheapfakes (WDC '23), July 10, 2023, Melbourne, VIC, Australia}
\acmDOI{10.1145/3595353.3595880}
\acmISBN{979-8-4007-0203-7/23/07}

\begin{document}

\title{Deepfake in the Metaverse: Security Implications for Virtual Gaming, Meetings, and Offices}

\author{Shahroz Tariq}
\affiliation{%
  \institution{CSIRO's Data61, Australia}
  \country{}}
\email{shahroz.tariq@data61.csiro.au}

\author{Alsharif Abuadbba}
\affiliation{%
  \institution{CSIRO's Data61, Australia}
  \country{}}
\email{sharif.abuadbba@data61.csiro.au}

\author{Kristen Moore}
\affiliation{%
  \institution{CSIRO's Data61, Australia}
  \country{}}
\email{kristen.moore@data61.csiro.au}





\begin{abstract}
The metaverse has gained significant attention from various industries due to its potential to create a fully immersive and interactive virtual world. However, the integration of deepfakes in the metaverse brings serious security implications, particularly with regard to impersonation. This paper examines the security implications of deepfakes in the metaverse, specifically in the context of gaming, online meetings, and virtual offices. The paper discusses how deepfakes can be used to impersonate in gaming scenarios, how online meetings in the metaverse open the door for impersonation, and how virtual offices in the metaverse lack physical authentication, making it easier for attackers to impersonate someone. The implications of these security concerns are discussed in relation to the confidentiality, integrity, and availability (CIA) triad. The paper further explores related issues such as the darkverse, and digital cloning, as well as regulatory and privacy concerns associated with addressing security threats in the virtual world.
\end{abstract}
\maketitle
\section{Introduction}

\begin{figure}
    \centering
\includegraphics[clip, trim=26.5pt 30.5pt 27pt 30pt, width=1\linewidth]{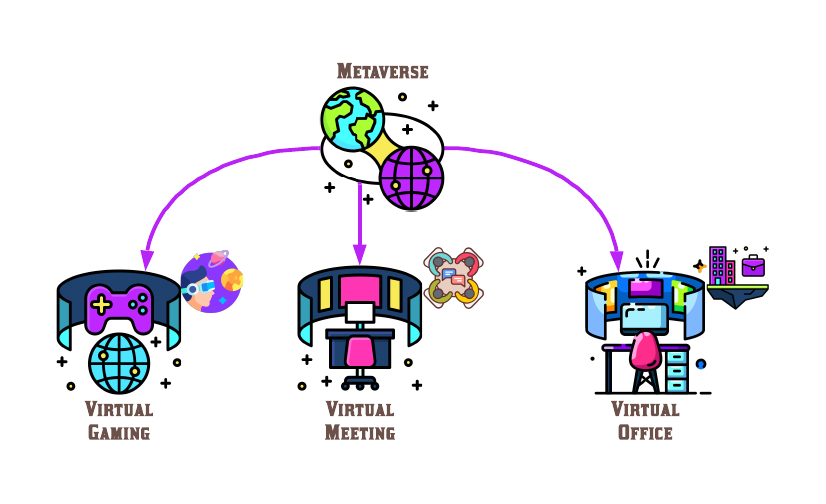}
    \caption{The three most commonly publicized scenarios in the metaverse: virtual gaming, virtual meetings, and virtual offices. These applications highlight the potential for immersive virtual experiences, but also raise concerns about security and privacy in this emerging technology.}
    \label{fig:scenarios}
\end{figure}













 The emergence of the metaverse~\cite{Metaverse_Survey} has captured the attention of the technology community, giving rise to widespread anticipation and debate. Prominent companies such as Meta (formerly Facebook), Microsoft, and Nvidia have expressed interest in the concept of a fully immersive virtual world, where individuals can interact with one another and their surroundings. To this end, various companies are releasing their own metaverse experiences, including Meta's Horizon Worlds~\cite{Facebook_metaverse,Meta_Horizon_Worlds}, Roblox's gaming metaverse~\cite{Roblox_metaverse}, Microsoft's Mesh~\cite{Microsoft_Mesh}, and Nvidia's Omniverse~\cite{NVIDIA_Omniverse}. While the potential applications of the metaverse are vast, with possibilities ranging from gaming to virtual meetings and offices, each method has its own benefits and limitations.

However, the applicability of deepfakes~\cite{Deepfake_survey,le2023deepfake} in the metaverse presents significant security implications, particularly with regard to impersonation. Deepfakes are computer-generated images or videos that can be manipulated to look like real people or events. The ability to generate such content has significantly increased with advances in machine learning and artificial intelligence. Recently, deepfakes in the metaverse have become a topic of discussion on different forums and media articles~\cite{Media_DeepfakesInMetaverse_1,Media_DeepfakesInMetaverse_2,Media_DeepfakesInMetaverse_3,Media_DeepfakesInMetaverse_4}.

The utilisation of metaverse technology has been associated with numerous benefits, including the provision of a fully immersive virtual environment that facilitates interaction with other users and virtual objects. Nevertheless, the potential deployment of deepfake technology to perpetrate malicious activities, such as impersonation, has shed light on the limitations of the technology and the need for effective security measures to be put in place. It is noteworthy that existing state-of-the-art deepfake detection methods~\cite{ShallowNet1,ShallowNet2,SamGAN,SamTAR,CLRNetold,CLRNet,MinhaFRETAL,MinhaCORED,ShahrozAWS,HasamFakeAVCeleb,HasamACMMM,JeonghoPTD} primarily focus on detecting deepfakes in the physical world and do not take into account the possibility of deepfakes in the metaverse. Thus, the current research is challenged with an evident gap in the identification and prevention of deepfakes in the metaverse, making it imperative to shed light on this research gap. 

In this paper, we explore the security implications of deepfakes in the metaverse. We will start by defining the concept of the metaverse and how it is expected to be used in three scenarios: (i) gaming, (ii) online meetings, and (iii) virtual offices (see Fig.~\ref{fig:scenarios}). We will also discuss the potential dangers of deepfake in each of the three scenarios by exploring the potential consequences of deepfake misuse in the metaverse, such as the ability to impersonate others, manipulate meetings, and disrupt virtual work environments. We will also discuss potential solutions to mitigate these risks and ensure the safety and security of metaverse users. We also explore the security implications of deepfakes in the metaverse to fake digital identity, the CIA triad, legal and regulatory challenges, privacy issues, and darkverse. Through this work, we aim to explore the potential security implications of deepfakes in the metaverse and to raise awareness of the risks and challenges posed by this technology.

\section{What is the Metaverse?}
The metaverse is a collective virtual shared space that offers an immersive and interactive 3D environment, facilitating real-time engagement with digital content and other individuals through advanced technologies such as virtual reality (VR) and augmented reality (AR). While originally a concept in science fiction, such as in Neal Stephenson's novel "Snow Crash" and the movie "The Matrix," recent technological advances have made the metaverse increasingly feasible. As a new form of social and economic infrastructure, the metaverse provides opportunities for people to work, play, socialize, learn, and consume content within a shared virtual space. While different visions of the metaverse exist among companies, organizations, and individuals, ranging from a fully autonomous world to a combination of different virtual platforms and experiences, the metaverse is considered a transformative technology that has the potential to  impact various aspects of our lives.

\section{Gaming in the Metaverse}
Gaming in the metaverse encompasses playing video games in a virtual world shared by millions globally, including massively multiplayer online games (MMOs), social games, and casual games such as puzzle and card games. These games offer interactive and detailed virtual worlds, providing players with opportunities to explore and interact with their surroundings.

The use of deepfakes in the context of gaming in the metaverse raises significant security concerns. Potential issues include identity theft, cyberbullying, distribution of malware, non-fungible token (NFT) scams, and intellectual property theft. As a significant demographic within metaverse gaming~\cite{Kids_metaverse_gaming}, minors are particularly vulnerable to these threats due to their limited experience and knowledge of online safety, which can result in sexual exploitation, social engineering, online grooming, and exposure to misinformation.

To address these risks, gaming companies must invest in advanced security measures, including identity verification systems, content monitoring and moderation tools, and anti-malware software. Additionally, players should be educated about the dangers of deepfakes and encouraged to report any suspicious activity encountered in the metaverse. Parents and guardians should take an active role in educating minors on the risks of deepfakes, monitoring their online activities, and encouraging them to report any questionable behavior. Online safety education, privacy settings, and parental controls can also help safeguard minors from the potential harms of deepfakes. 

\section{Online Meetings in the Metaverse}
Online meetings in the metaverse offer a virtual space for individuals to communicate and collaborate within a shared immersive digital environment, spanning from basic text-based chat rooms to fully immersive 3D environments, and utilizing voice chat, instant messaging, or other means of communication. These virtual meetings are versatile and can serve various purposes, such as team collaboration, networking, socializing, or attending virtual events, such as conferences. 

However, the potential risks to privacy and reputation are significant due to the metaverse's ability to create an opportunity for attackers to impersonate others. Deepfake technology can be utilized to deceive others through impersonation, leading to potential fraud or espionage, as demonstrated in a recent example of Elon Musk's deepfake zoom-bombing online meetings~\cite{Elon_ZoomBomb}. In addition, the authenticity and trust of participants may be compromised by the creation of convincing deepfakes, which could undermine trust and collaboration.

To address these issues, it is essential to implement measures such as identity verification, digital signatures, or other security measures to ensure the authenticity of participants. Moreover, the development of tools and technologies that can detect various forms of deepfakes in the metaverse and prevent their use during online meetings may be necessary. Overall, while deepfake technology poses a disruptive risk to online meetings in the metaverse, it is also possible to mitigate these risks effectively by utilizing appropriate security measures and technology.

\section{Virtual Offices in the Metaverse}

The emergence of virtual offices or workplaces in the metaverse is a recent but promising development that has the potential to revolutionize collaboration and work practices. By leveraging the metaverse's virtual environment, colleagues can work together in a customizable, shared digital workspace that offers numerous benefits, such as reduced overhead cost, increased flexibility, and access to a global talent pool. In addition, virtual offices in the metaverse can enable more dynamic and immersive meetings, greater collaboration, and increased creativity.

However, the virtual nature of the metaverse presents a security challenge, as attackers can impersonate team members through the use of deepfakes, leading to data breaches and financial loss. For instance, deepfakes can be employed to create fake identities or to impersonate colleagues, which could lead to trust issues and confusion within the team. For example, an employee could use a deepfake to create a fake version of their boss or co-worker to make it appear as if they are giving instructions. Furthermore, deepfakes can be used to spread false information or propaganda, potentially impacting important decisions.

To address the potential threats posed by deepfakes in virtual offices or workplaces in the metaverse, clear guidelines and protocols are required to verify team members' identities and the authenticity of content shared in the virtual environment. This can be critical in establishing one's innocence in the case of a crime. It is also important to remain abreast of the latest deepfake technology developments and to leverage tools and software that can aid in identifying and detecting deepfakes.

\section{Discussion}


\noindent
\textbf{Fake Digital Identity and Cloning in the digital world. } 
One of the central ideas underlying the concept of the metaverse is the ability for individuals to create digital replicas of themselves, known as avatars, in the virtual world. These avatars are designed to mimic the physical appearance and behavior of their real-life counterparts, allowing individuals to interact with one another in the digital realm. However, the ability to clone oneself in the metaverse also raises concerns about the potential for impersonation. Unlike the physical world, where impersonating someone convincingly is challenging, it is much easier to create a convincing digital clone of a person in the metaverse due to the abundance of personal information available on the internet that can be used to create deepfakes. The possibility of an attacker using deepfakes to impersonate someone in the metaverse is a significant concern, as it could be used to commit various illicit activities. 
One potential solution to this problem is the implementation of digital identity verification systems. Such systems could use biometric data, such as facial recognition, to verify an individual's identity before allowing them to create a digital avatar. By doing so, attackers would be prevented from creating digital clones of other people without their consent, thereby ensuring a higher level of security in the metaverse.

\noindent
\textbf{Deepfakes impact on CIA Triad in the Metaverse. } The CIA Triad, established by the National Institute of Standards and Technology (NIST), comprises confidentiality, integrity, and availability, which are the three primary objectives of information security. Confidentiality safeguards sensitive information by ensuring that only authorized parties can access it. Integrity ensures that information remains accurate and unaltered, while availability guarantees that authorized parties can access information when necessary. In the context of the metaverse, deepfakes pose a potential threat to the CIA Triad's objectives of confidentiality, integrity, and availability. Specifically, deepfakes have the potential to compromise confidentiality by enabling the impersonation of authorized individuals, thereby permitting unauthorized access to sensitive areas. Furthermore, deepfakes can undermine the integrity of information, images, or videos, by spreading false or misleading information about individuals or organizations, causing reputational harm. Finally, deepfakes can also disrupt availability by disseminating propaganda or fake news, leading to confusion and chaos. 




\noindent
\textbf{Legal and Regulatory Challenges. } The lack of regulations regarding the application of laws from the physical world in the metaverse presents a significant challenge. For instance, identifying and prosecuting an offender who has committed a crime in the virtual world using deepfake impersonation can be difficult. Additionally, jurisdictional issues arise due to the existence of varying laws in different countries. In the event that an attacker is located in a country where there are no legal consequences for their actions in the metaverse, holding them accountable can become problematic. Consequently, a universal set of rules and regulations for the metaverse becomes difficult to establish, given that different countries may have different interpretations of what constitutes criminal behavior. To address security concerns in the metaverse, a coordinated effort between governments, regulatory bodies, and technology companies is necessary~\cite{Deepfakes_In_Metaverse_Legal_Issues}. This entails the development of universally applicable standards and regulations that can transcend geographical and jurisdictional barriers. Also, continuous efforts toward the development of new technologies capable of preventing and detecting criminal activities in the metaverse are also required. 

\noindent
\textbf{Privacy issues. } Although digital identity verification systems can serve as a potential measure for mitigating deepfake-based impersonation in the metaverse, the utilization of these systems raises concerns. The apprehension stems from the possibility of digital identity verification systems being utilized to track and monitor individuals' virtual activities, consequently, potentially infringing on their privacy and freedom. It is argued that any digital identity verification system deployed in the metaverse must maintain a balance between security and the need for privacy and freedom.



\noindent
\textbf{Darkverse. }
The rise of metaverse technology has created new opportunities for both legitimate users and malicious actors. One of the primary concerns is the creation of private spaces that enable illegal activities and communication among criminals, which Trend Micro refers to as the \textit{darkverse}~\cite{darkverse}. This space operates similarly to the dark web, but it exists within the metaverse and is unindexed, making it challenging to locate via standard search engines. The darkverse's pseudo-physical user presence makes it more dangerous than the dark web, as criminals can use proximity-based messaging or other methods to conceal their communications, rendering them difficult for law enforcement agencies to intercept. Darkverse could be used to facilitate illegal activities such as deepfake-based revenge pornography and misinformation campaigns. 
Despite the possibility of the darkverse being a space for free speech, the primary objective of these spaces is to facilitate illegal activities, and it may become a safe haven for criminals seeking to engage in such activities with minimal risk of detection.

\section{Conclusion}
In conclusion, deepfakes in the metaverse present significant security implications, particularly around impersonation. The three scenarios of gaming, online meetings, and virtual offices serve as examples of how these security implications can play out in practice. The lack of physical authentication in the metaverse makes it easier for attackers to impersonate others and commit crimes without being held accountable. Mitigating these security implications will require a combination of technological solutions and legal frameworks that balance security and privacy concerns. As the metaverse continues to evolve, it is important to address these issues proactively to ensure a safe and secure virtual environment for all users.
\begin{acks}
This research was financially supported by CSIRO's Collaborative Intelligence Future Science Platform. The diagram has been designed using images from Flaticon.com
\end{acks}
\bibliographystyle{ACM-Reference-Format}
\bibliography{ref}


\begin{thebibliography}{28}


\ifx \showCODEN    \undefined \def \showCODEN     #1{\unskip}     \fi
\ifx \showDOI      \undefined \def \showDOI       #1{#1}\fi
\ifx \showISBNx    \undefined \def \showISBNx     #1{\unskip}     \fi
\ifx \showISBNxiii \undefined \def \showISBNxiii  #1{\unskip}     \fi
\ifx \showISSN     \undefined \def \showISSN      #1{\unskip}     \fi
\ifx \showLCCN     \undefined \def \showLCCN      #1{\unskip}     \fi
\ifx \shownote     \undefined \def \shownote      #1{#1}          \fi
\ifx \showarticletitle \undefined \def \showarticletitle #1{#1}   \fi
\ifx \showURL      \undefined \def \showURL       {\relax}        \fi
\providecommand\bibfield[2]{#2}
\providecommand\bibinfo[2]{#2}
\providecommand\natexlab[1]{#1}
\providecommand\showeprint[2][]{arXiv:#2}

\bibitem[Cole(2020)]%
        {Elon_ZoomBomb}
\bibfield{author}{\bibinfo{person}{Samantha Cole}.}
  \bibinfo{year}{2020}\natexlab{}.
\newblock \bibinfo{title}{This Open-Source Program Deepfakes You During Zoom
  Meetings, in Real Time}.
\newblock
\newblock
\urldef\tempurl%
\url{https://www.vice.com/en/article/g5xagy/this-open-source-program-deepfakes-you-during-zoom-meetings-in-real-time}
\showURL{%
\tempurl}
\newblock
\shownote{Accessed: 13-March-2023}.


\bibitem[del Castillo(2022)]%
        {Media_DeepfakesInMetaverse_2}
\bibfield{author}{\bibinfo{person}{Michael del Castillo}.}
  \bibinfo{year}{2022}\natexlab{}.
\newblock \bibinfo{title}{Facebook's Metaverse Could Be Overrun By Deep Fakes
  And Other Misinformation If These Non-Profits Don’t Succeed}.
\newblock
\newblock
\urldef\tempurl%
\url{https://www.forbes.com/sites/michaeldelcastillo/2022/08/29/facebooks-metaverse-could-be-overrun-by-deep-fakes-and-other-misinformation-if-these-non-profits-dont-succeed/?sh=185318742737}
\showURL{%
\tempurl}
\newblock
\shownote{Accessed: 13-March-2023}.


\bibitem[Khalid et~al\mbox{.}(2021a)]%
        {HasamACMMM}
\bibfield{author}{\bibinfo{person}{Hasam Khalid}, \bibinfo{person}{Minha Kim},
  \bibinfo{person}{Shahroz Tariq}, {and} \bibinfo{person}{Simon~S Woo}.}
  \bibinfo{year}{2021}\natexlab{a}.
\newblock \showarticletitle{Evaluation of an Audio-Video Multimodal Deepfake
  Dataset using Unimodal and Multimodal Detectors}. In
  \bibinfo{booktitle}{\emph{Proceedings of the 1st Workshop on Synthetic
  Multimedia-Audiovisual Deepfake Generation and Detection}}.
  \bibinfo{pages}{7--15}.
\newblock


\bibitem[Khalid et~al\mbox{.}(2021b)]%
        {HasamFakeAVCeleb}
\bibfield{author}{\bibinfo{person}{Hasam Khalid}, \bibinfo{person}{Shahroz
  Tariq}, \bibinfo{person}{Minha Kim}, {and} \bibinfo{person}{Simon~S Woo}.}
  \bibinfo{year}{2021}\natexlab{b}.
\newblock \showarticletitle{FakeAVCeleb: A Novel Audio-Video Multimodal
  Deepfake Dataset}.
\newblock \bibinfo{journal}{\emph{arXiv preprint arXiv:2108.05080}}
  (\bibinfo{year}{2021}).
\newblock


\bibitem[Kim et~al\mbox{.}(2022)]%
        {JeonghoPTD}
\bibfield{author}{\bibinfo{person}{Jeongho Kim}, \bibinfo{person}{Shahroz
  Tariq}, {and} \bibinfo{person}{Simon~S Woo}.}
  \bibinfo{year}{2022}\natexlab{}.
\newblock \showarticletitle{PTD: Privacy-Preserving Human Face Processing
  Framework using Tensor Decomposition}. In
  \bibinfo{booktitle}{\emph{Proceedings of the 37th ACM/SIGAPP Symposium on
  Applied Computing}}. \bibinfo{pages}{1296--1303}.
\newblock
\urldef\tempurl%
\url{https://doi.org/10.1145/3477314.3507036}
\showDOI{\tempurl}


\bibitem[Kim et~al\mbox{.}(2021a)]%
        {MinhaCORED}
\bibfield{author}{\bibinfo{person}{Minha Kim}, \bibinfo{person}{Shahroz Tariq},
  {and} \bibinfo{person}{Simon~S Woo}.} \bibinfo{year}{2021}\natexlab{a}.
\newblock \showarticletitle{Cored: Generalizing fake media detection with
  continual representation using distillation}. In
  \bibinfo{booktitle}{\emph{Proceedings of the 29th ACM International
  Conference on Multimedia}}. \bibinfo{pages}{337--346}.
\newblock


\bibitem[Kim et~al\mbox{.}(2021b)]%
        {MinhaFRETAL}
\bibfield{author}{\bibinfo{person}{Minha Kim}, \bibinfo{person}{Shahroz Tariq},
  {and} \bibinfo{person}{Simon~S Woo}.} \bibinfo{year}{2021}\natexlab{b}.
\newblock \showarticletitle{FReTAL: Generalizing Deepfake Detection using
  Knowledge Distillation and Representation Learning}. In
  \bibinfo{booktitle}{\emph{Proceedings of the IEEE/CVF Conference on Computer
  Vision and Pattern Recognition}}. \bibinfo{pages}{1001--1012}.
\newblock


\bibitem[Kleeman(2021)]%
        {Kids_metaverse_gaming}
\bibfield{author}{\bibinfo{person}{David Kleeman}.}
  \bibinfo{year}{2021}\natexlab{}.
\newblock \bibinfo{title}{Kids have Kickstarted the Metaverse}.
\newblock
\newblock
\urldef\tempurl%
\url{https://techonomy.com/kids-have-kickstarted-the-metaverse/}
\showURL{%
\tempurl}
\newblock
\shownote{Accessed: 13-March-2023}.


\bibitem[Le et~al\mbox{.}(2023)]%
        {le2023deepfake}
\bibfield{author}{\bibinfo{person}{Binh Le}, \bibinfo{person}{Shahroz Tariq},
  \bibinfo{person}{Alsharif Abuadbba}, \bibinfo{person}{Kristen Moore}, {and}
  \bibinfo{person}{Simon Woo}.} \bibinfo{year}{2023}\natexlab{}.
\newblock \showarticletitle{Why Do Deepfake Detectors Fail?}
\newblock \bibinfo{journal}{\emph{arXiv preprint arXiv:2302.13156}}
  (\bibinfo{year}{2023}).
\newblock


\bibitem[Lee et~al\mbox{.}(2021a)]%
        {SamTAR}
\bibfield{author}{\bibinfo{person}{Sangyup Lee}, \bibinfo{person}{Shahroz
  Tariq}, \bibinfo{person}{Junyaup Kim}, {and} \bibinfo{person}{Simon~S Woo}.}
  \bibinfo{year}{2021}\natexlab{a}.
\newblock \showarticletitle{TAR: Generalized Forensic Framework to Detect
  Deepfakes Using Weakly Supervised Learning}. In
  \bibinfo{booktitle}{\emph{IFIP International Conference on ICT Systems
  Security and Privacy Protection}}. Springer, \bibinfo{pages}{351--366}.
\newblock


\bibitem[Lee et~al\mbox{.}(2021b)]%
        {SamGAN}
\bibfield{author}{\bibinfo{person}{Sangyup Lee}, \bibinfo{person}{Shahroz
  Tariq}, \bibinfo{person}{Youjin Shin}, {and} \bibinfo{person}{Simon~S Woo}.}
  \bibinfo{year}{2021}\natexlab{b}.
\newblock \showarticletitle{Detecting handcrafted facial image manipulations
  and GAN-generated facial images using Shallow-FakeFaceNet}.
\newblock \bibinfo{journal}{\emph{Applied Soft Computing}}
  \bibinfo{volume}{105} (\bibinfo{year}{2021}), \bibinfo{pages}{107256}.
\newblock


\bibitem[Levy(2022)]%
        {Media_DeepfakesInMetaverse_4}
\bibfield{author}{\bibinfo{person}{Steven Levy}.}
  \bibinfo{year}{2022}\natexlab{}.
\newblock \bibinfo{title}{What’s Deepfake Bruce Willis Doing in My
  Metaverse?}
\newblock
\newblock
\urldef\tempurl%
\url{https://www.wired.com/story/plaintext-bruce-willis-deepfake-metaverse}
\showURL{%
\tempurl}
\newblock
\shownote{Accessed: 13-March-2023}.


\bibitem[LLP(2022)]%
        {Deepfakes_In_Metaverse_Legal_Issues}
\bibfield{author}{\bibinfo{person}{Reed~Smith LLP}.}
  \bibinfo{year}{2022}\natexlab{}.
\newblock \bibinfo{booktitle}{\emph{Reed Smith Guide to the Metaverse, 2nd
  Edition}}.
\newblock \bibinfo{publisher}{Reed Smith LLP}.
\newblock
\urldef\tempurl%
\url{https://www.reedsmith.com/en/perspectives/metaverse}
\showURL{%
\tempurl}


\bibitem[Micro(2023)]%
        {darkverse}
\bibfield{author}{\bibinfo{person}{Trend Micro}.}
  \bibinfo{year}{2023}\natexlab{}.
\newblock \bibinfo{title}{Darkverse}.
\newblock
\newblock
\urldef\tempurl%
\url{https://www.trendmicro.com/vinfo/us/security/definition/darkverse}
\showURL{%
\tempurl}
\newblock
\shownote{Accessed: 13-March-2023}.


\bibitem[Microsoft(2023)]%
        {Microsoft_Mesh}
\bibfield{author}{\bibinfo{person}{Microsoft}.}
  \bibinfo{year}{2023}\natexlab{}.
\newblock \bibinfo{title}{Microsoft Mesh}.
\newblock
\newblock
\urldef\tempurl%
\url{https://www.microsoft.com/en-us/mesh}
\showURL{%
\tempurl}
\newblock
\shownote{Accessed: 13-March-2023}.


\bibitem[Mirsky and Lee(2021)]%
        {Deepfake_survey}
\bibfield{author}{\bibinfo{person}{Yisroel Mirsky} {and} \bibinfo{person}{Wenke
  Lee}.} \bibinfo{year}{2021}\natexlab{}.
\newblock \showarticletitle{The creation and detection of deepfakes: A survey}.
\newblock \bibinfo{journal}{\emph{ACM Computing Surveys (CSUR)}}
  \bibinfo{volume}{54}, \bibinfo{number}{1} (\bibinfo{year}{2021}),
  \bibinfo{pages}{1--41}.
\newblock


\bibitem[Nvidia(2023)]%
        {NVIDIA_Omniverse}
\bibfield{author}{\bibinfo{person}{Nvidia}.} \bibinfo{year}{2023}\natexlab{}.
\newblock \bibinfo{title}{NVIDIA Omniverse}.
\newblock
\newblock
\urldef\tempurl%
\url{https://www.nvidia.com/en-gb/omniverse/}
\showURL{%
\tempurl}
\newblock
\shownote{Accessed: 13-March-2023}.


\bibitem[Platforms(2023a)]%
        {Facebook_metaverse}
\bibfield{author}{\bibinfo{person}{Meta Platforms}.}
  \bibinfo{year}{2023}\natexlab{a}.
\newblock \bibinfo{title}{Facebook Metaverse}.
\newblock
\newblock
\urldef\tempurl%
\url{https://about.meta.com/metaverse}
\showURL{%
\tempurl}
\newblock
\shownote{Accessed: 13-March-2023}.


\bibitem[Platforms(2023b)]%
        {Meta_Horizon_Worlds}
\bibfield{author}{\bibinfo{person}{Meta Platforms}.}
  \bibinfo{year}{2023}\natexlab{b}.
\newblock \bibinfo{title}{Meta Horizon Worlds}.
\newblock
\newblock
\urldef\tempurl%
\url{https://www.meta.com/gb/en/horizon-worlds/}
\showURL{%
\tempurl}
\newblock
\shownote{Accessed: 13-March-2023}.


\bibitem[Raturi(2022)]%
        {Roblox_metaverse}
\bibfield{author}{\bibinfo{person}{Gautam Raturi}.}
  \bibinfo{year}{2022}\natexlab{}.
\newblock \bibinfo{title}{Roblox Metaverse: Everything You Need to Know}.
\newblock
\newblock
\urldef\tempurl%
\url{https://medium.com/codex/everything-you-need-to-know-about-the-roblox-metaverse-928e9531e693}
\showURL{%
\tempurl}
\newblock
\shownote{Accessed: 13-March-2023}.


\bibitem[Stewart(2021)]%
        {Media_DeepfakesInMetaverse_3}
\bibfield{author}{\bibinfo{person}{Darin Stewart}.}
  \bibinfo{year}{2021}\natexlab{}.
\newblock \bibinfo{title}{Maverick Research: Deepfakes Will Kill the Metaverse;
  Synthetic Media Could Save It}.
\newblock
\newblock
\urldef\tempurl%
\url{https://www.gartner.com/en/documents/4008295}
\showURL{%
\tempurl}
\newblock
\shownote{Accessed: 13-March-2023}.


\bibitem[Tariq et~al\mbox{.}(2021a)]%
        {ShahrozAWS}
\bibfield{author}{\bibinfo{person}{Shahroz Tariq}, \bibinfo{person}{Sowon
  Jeon}, {and} \bibinfo{person}{Simon Woo}.} \bibinfo{year}{2021}\natexlab{a}.
\newblock \showarticletitle{Am I a Real or Fake Celebrity? Measuring Commercial
  Face Recognition Web APIs under Deepfake Impersonation Attack}.
\newblock \bibinfo{journal}{\emph{arXiv preprint arXiv:2103.00847}}
  (\bibinfo{year}{2021}).
\newblock


\bibitem[Tariq et~al\mbox{.}(2018)]%
        {ShallowNet1}
\bibfield{author}{\bibinfo{person}{Shahroz Tariq}, \bibinfo{person}{Sangyup
  Lee}, \bibinfo{person}{Hoyoung Kim}, \bibinfo{person}{Youjin Shin}, {and}
  \bibinfo{person}{Simon~S Woo}.} \bibinfo{year}{2018}\natexlab{}.
\newblock \showarticletitle{Detecting both machine and human created fake face
  images in the wild}. In \bibinfo{booktitle}{\emph{Proceedings of the 2nd
  International Workshop on Multimedia Privacy and Security}}. ACM,
  \bibinfo{pages}{81--87}.
\newblock


\bibitem[Tariq et~al\mbox{.}(2019)]%
        {ShallowNet2}
\bibfield{author}{\bibinfo{person}{Shahroz Tariq}, \bibinfo{person}{Sangyup
  Lee}, \bibinfo{person}{Hoyoung Kim}, \bibinfo{person}{Youjin Shin}, {and}
  \bibinfo{person}{Simon~S Woo}.} \bibinfo{year}{2019}\natexlab{}.
\newblock \showarticletitle{GAN is a friend or foe?: a framework to detect
  various fake face images}. In \bibinfo{booktitle}{\emph{Proceedings of the
  34th ACM/SIGAPP Symposium on Applied Computing}}. ACM,
  \bibinfo{pages}{1296--1303}.
\newblock


\bibitem[Tariq et~al\mbox{.}(2021b)]%
        {CLRNet}
\bibfield{author}{\bibinfo{person}{Shahroz Tariq}, \bibinfo{person}{Sangyup
  Lee}, {and} \bibinfo{person}{Simon Woo}.} \bibinfo{year}{2021}\natexlab{b}.
\newblock \showarticletitle{One detector to rule them all: Towards a general
  deepfake attack detection framework}. In
  \bibinfo{booktitle}{\emph{Proceedings of the web conference 2021}}.
  \bibinfo{pages}{3625--3637}.
\newblock


\bibitem[Tariq et~al\mbox{.}(2020)]%
        {CLRNetold}
\bibfield{author}{\bibinfo{person}{Shahroz Tariq}, \bibinfo{person}{Sangyup
  Lee}, {and} \bibinfo{person}{Simon~S Woo}.} \bibinfo{year}{2020}\natexlab{}.
\newblock \showarticletitle{A Convolutional LSTM based Residual Network for
  Deepfake Video Detection}.
\newblock \bibinfo{journal}{\emph{arXiv preprint arXiv:2009.07480}}
  (\bibinfo{year}{2020}).
\newblock


\bibitem[Wang et~al\mbox{.}(2022)]%
        {Metaverse_Survey}
\bibfield{author}{\bibinfo{person}{Yuntao Wang}, \bibinfo{person}{Zhou Su},
  \bibinfo{person}{Ning Zhang}, \bibinfo{person}{Rui Xing},
  \bibinfo{person}{Dongxiao Liu}, \bibinfo{person}{Tom~H Luan}, {and}
  \bibinfo{person}{Xuemin Shen}.} \bibinfo{year}{2022}\natexlab{}.
\newblock \showarticletitle{A survey on metaverse: Fundamentals, security, and
  privacy}.
\newblock \bibinfo{journal}{\emph{IEEE Communications Surveys \& Tutorials}}
  (\bibinfo{year}{2022}).
\newblock


\bibitem[Woollacott(2022)]%
        {Media_DeepfakesInMetaverse_1}
\bibfield{author}{\bibinfo{person}{Emma Woollacott}.}
  \bibinfo{year}{2022}\natexlab{}.
\newblock \bibinfo{title}{Rise of deepfakes: who can you trust in the
  metaverse?}
\newblock
\newblock
\urldef\tempurl%
\url{https://cybernews.com/security/rise-of-deepfakes}
\showURL{%
\tempurl}
\newblock
\shownote{Accessed: 13-March-2023}.


\end{thebibliography}
\end{document}